\documentstyle[multicol,prb,aps]{revtex}
\begin{document}

\draft

\title{Kondo-resonance, Coulomb blockade, and Andreev transport through a
       quantum dot }

\author{Kicheon Kang\cite{kang}}

\address{   Max-Planck-Institut f\"ur Physik Komplexer Systeme,
            N\"othnitzer Str. 38, D-01187 Dresden, Germany }

\date{\today}

\maketitle

\begin{abstract}
We study resonant tunneling through an interacting quantum dot
coupled to normal metallic and superconducting leads. We show that
large Coulomb interaction gives rise to novel effects in Andreev transport.
Adopting an exact relation for the Green's function, we find
that at zero temperature, the linear response conductance is enhanced due to
Kondo-Andreev resonance in the Kondo limit, while it is suppressed
in the empty site limit.  In the Coulomb blockaded region,
on the other hand, the conductance is reduced more than the corresponding
conductance with normal leads 
because large charging energy suppresses Andreev reflection.
\end{abstract}
%
\begin{multicols}{2}
Electronic transport of mesoscopic devices containing superconducting 
electrode has been an interesting subject in recent 
years.~\cite{lambert98} Transmission of electrons through normal metal -
superconductor (N-S) interfaces requires the conversion of normal 
current to supercurrent, which is called Andreev reflection.~\cite{andreev64}
With the recent advances of nanofabrication technique, quantum 
interference effects have been extensively studied in the mesoscopic
N-S heterostructures (see e.g. references in \cite{lambert98}). 
In a phase coherent N-S structures, the phase of quasi-particles
as well as Cooper pairs is preserved and transport properties depend
strongly on the nature of quasiparticle phase.
Theoretically, Landauer-B\"uttiker type
formula~\cite{landauer70,buttiker86} has been used extensively to
describe quantum transport in many kind of N-S hybrid structures
using non-interacting models. (For review, see e.g. 
\cite{lambert98,beenakker97}.) 

Resonant tunneling through an interacting quantum dot (QD) or Anderson
impurity has been intensively investigated recently. It has been
shown that large Coulomb
interaction gives rise to anomalous properties in transport. An example
is the Kondo-resonant transport. Kondo-resonant
transport has been predicted theoretically~\cite{ng88,hershfeld,meir,yeyati93}
and verified experimentally~\cite{ralph94,gordon97,cronen98}
by conductance measurements for artificially made Anderson 
impurities.
On the other side, strong electron-electron interactions
suppress conductance peaks 
where the systems are weakly coupled to the leads.
It has been shown that the electron-electron interactions 
lead to conductance suppression due to the orthogonality 
catastrophe.~\cite{kinaret92,matveev-golden96,stafford98}
Stafford {\it et al.}~\cite{stafford98}
showed that the coherent transmission in artificial molecule structures
is suppressed with increasing the system size by using the Hubbard-type
model. 
Coulomb interaction has been found to play a crucial role 
in the nature of the transmission phase.~\cite{bruder96,oreg97,kang98}
It has been shown that Coulomb interactions give rise to anomalous effects
in phase evolution through  a quantum dot embedded in an arm
of the Aharonov-Bohm interferometer,
such as an inter-resonance phase drop.
Non-equilibrium transport in an interacting quantum dot 
where both leads are superconductors
has been studied recently by using the non-equilibrium Green's function
method.~\cite{yeyati97,kang98a} Andreev reflection has been supposed
to be negligible in the weak tunneling limit because large
charging energy leads to Coulomb blockade of Andreev transport. 
In the meanwhile, for
a moderately coupled quantum dot,
multiple Andreev
reflections give rise to a novel
subgap structure in the current-voltage
curve due to resonant tunneling,
which are quite different from those of S-S contacts.~\cite{yeyati97}
Resonant Andreev tunneling in strongly correlated quantum dot coupled to
normal and superconducting leads has been investigated recently
by Fazio and Raimondi.~\cite{fazio98} Using the non-equilibrium Green's
function formalism and equation of motion technique 
they have shown that the Kondo-resonant transmission is 
enhanced in the limit of large Coulomb repulsion
due to the existence of a superconducting electrode. 

In this paper, we investigate coherent 
transport through an interacting quantum
dot coupled to normal and superconducting electrodes based on the
scattering matrix formulation. We consider 
a model as shown schematically in Fig.1, where normal scattering and
Andreev reflection are decoupled. In the QD-${\rm N}_2$ boundary, only
normal scattering is taken into account while Andreev reflection
is considered in the ${\rm N}_2$-S boundary. It is assumed that the
${\rm N}_2$-S boundary is perfect and the normal scattering doesn't
occur at this boundary. This model is applicable
to microjunctions where the length scale of normal scattering 
and Andreev reflection is
well separated.~\cite{beenakker92} With this model, Landauer
type formula for the linear response conductance has been derived by 
Beenakker~\cite{beenakker92}
in the framework of non-interacting electron model.
In the presence of interactions, this formula
cannot be used in general because of the presence of inelastic
processes. However, in the linear response regime ($V=0$) with
zero temperature, there is no phase space for
inelastic processes and the formula can
be equally applied to the system containing interactions.
The linear response conductance for the system under consideration
can be written as~\cite{beenakker92}:
\begin{equation}
 G_{NS} = \frac{4e^2}{h} \sum_n\frac{ T_n^2 }{ (2-T_n)^2 } ,
\end{equation}
where $T_n$ is transmission probability of $n$th channel. This equation is
valid in the absence of an applied magnetic field.
We consider single channel case where the transmission probability
through the quantum dot
is represented by $T_{QD}$ :
\begin{equation}
 G_{NS} = \frac{4e^2}{h} \frac{ T_{QD}^2 }{ (2-T_{QD})^2 } .
\end{equation}
The corresponding formula for the normal leads is the well
known Landauer formula~\cite{landauer70}
\begin{equation}
 G_N = \frac{2e^2}{h} T_{QD} .
\end{equation}

Let's consider an Anderson impurity for the quantum dot
with doubly degenerate level energy $\varepsilon_0$ and on-site
Coulomb repulsion strength $U\sim e^2/C$, $C$ being capacitance
of the dot. At zero temperature the transmission probability $T_{QD}$
can be obtained as follows owing to the fact that there
are no inelastic processes.~\cite{langreth66,ng88} 
Due to the absence of the inelastic
scattering, the imaginary part of the self-energy
for the Green's function at the Fermi energy $\varepsilon_F$ is given by
\begin{equation}
 \mbox{Im\,}\Sigma(\varepsilon_F) = -\Gamma/2  ,
   \label{eq:sigma}
\end{equation}
where $\Gamma=\Gamma_L+\Gamma_R$ and 
$\Gamma_L/\hbar$ and $\Gamma_R/\hbar$
are tunneling rate through left and right leads, respectively.
With this condition the average occupation on the dot can be written as
\begin{equation}
 \langle n \rangle = \frac{2}{\pi} \mbox{Im\,}
   [\log{G^r(\varepsilon_F)} ] . \label{eq:occ}
\end{equation}
At zero temperature, the transmission probability can be 
expressed in terms of the
exact Green's function as
\begin{equation}
 T_{QD} = \Gamma_L\Gamma_R |G^r(\varepsilon_F)|^2  ,
  \label{eq:TQD}
\end{equation}
which leads to the final expression with the help of the 
Eq.(\ref{eq:sigma}) 
\begin{equation}
 T_{QD} = \frac{4\Gamma_L\Gamma_R}{\Gamma^2} \sin^2{\varphi}
\end{equation}
where $\varphi=\pi\langle n\rangle/2$. ($\langle n\rangle$ is
average occupation of the dot.)

Here $\langle n\rangle$ is calculated numerically by an 
equation of motion method,~\cite{kang95} which has been shown to be
quite accurate for large $U$. 
We consider a symmetric coupling of the quantum dot to leads,
that is $\Gamma_L=\Gamma_R$.
From the calculated values of the average occupation, 
we display the conductances in the 
Fig.2 as a function
of $\varepsilon_F-\varepsilon_0$. 
The parameters used for calculations 
are $U=50\Gamma$ and $W=200\Gamma$, with $W$ being the bandwidth of
the leads. Since the transmission probability reaches one 
for $\varepsilon_F-\varepsilon_0 >> \Gamma$, the conductance of 
normal-superconductor
hybrid system goes to $4e^2/h$, which is twice of normal conductance.
This is a result of perfect transmission through the quantum dot
in the Kondo limit. On the contrary, the conductances are suppressed
in the empty site limit because of small transparency.
$G_{NS}$ decays faster than $G_N$ because transmission by
Andreev reflection
requires two particle tunneling through the quantum dot.

In real systems, nearly perfect Kondo-resonant 
transmission could not be realized 
though it is predicted by an exact relation at zero temperature.
When the temperature is larger than the Kondo temperature,
Kondo-resonant transmission does not occur and the
transport is suppressed by Coulomb repulsion,
rather than enhanced. 
In the case $k_BT << \Gamma$, thermal broadening
can be neglected and the conductances can be obtained
in the following way with an approximate Green's function.
If Kondo-like correlation is
neglected, the transparency can be obtained by an approximate
retarded Green's function~\cite{haug96} 
which is similar to the Breit-Wigner type
\begin{equation}
 G^r(\varepsilon) = \frac{ 1-\langle n\rangle/2 }{
   \varepsilon-\varepsilon_0+i\Gamma/2 }  +
   \frac{ \langle n\rangle/2 }{ \varepsilon-\varepsilon_0-U
     +i\Gamma/2 } . \label{eq:green}
\end{equation}
Note that this equation coincides with the Breit-Wigner formula
for $U=0$.
The self-consistent value of $\langle n\rangle$ 
is given by the relation
\begin{equation}
 \langle n\rangle = -\frac{2}{\pi} \int_{-\infty}^{\varepsilon_F}
  \mbox{Im}\, G^r(\varepsilon)\,d\varepsilon ,
\end{equation}
which leads to the expression~\cite{kang98}
\begin{equation}
  \langle n \rangle = \frac{ 1+2P_1 }{ 1+P_1-P_2 } ,
\end{equation}
where
\begin{eqnarray*}
 P_1 &=& \frac{1}{\pi} \arctan{ \frac{2(\varepsilon_F-\varepsilon_0)}{\Gamma} }
         ,  \\
 P_2 &=& \frac{1}{\pi} \arctan{ \frac{2(\varepsilon_F-\varepsilon_0-U)}{\Gamma}
}
   .
\end{eqnarray*}
Then we can get the conductance through the Eq.(\ref{eq:TQD})

Fig.3 displays the conductances obtained by the Eq.(\ref{eq:TQD}) and 
Eq.(\ref{eq:green}). As one can see, $G_{NS}$ is suppressed more than
$G_N$ even in the ``resonance" point. This phenomenon arises because
Coulomb interactions in the dot suppress coherent transmission through
the quantum dot. This would become a very general feature in transport
through interacting system as far as the coupling to the leads
are not so strong. If we consider larger system like a coupled chain of
quantum dots, transmission probability will be more reduced due to 
orthogonality catastrophe. So one could say that in general
$G_{NS}$ will be negligible compared
to $G_N$ in strongly interacting systems weakly coupled to leads. 
Normal conductance 
suppression with increasing the system
size has been studied by Stafford et al.~\cite{stafford98}
While the normal conductance is proportional to the transparency,
$G_{NS}$ is second order of transmission probability. So $G_{NS}$
will decrease faster than $G_N$ with increasing the system
size.  Even in the case of single quantum dot, we could see 
suppression of transmission due to the Coulomb interaction
from our calculations.

For comparison, we plot the conductances of non-interacting case
($U=0$) in the Fig.4. As well known from the Breit-Wigner formula, 
one can see that $G_{NS}=
2G_N$ in the resonance point because of perfect transmission.
Comparing Fig.3 and Fig.4, one can conclude that the large charging
energy suppress Andreev reflection even on resonance.  

In conclusion, we have discussed resonant tunneling through a
strongly interacting quantum dot
coupled to normal metallic and superconducting leads. We have found
that in strongly interacting quantum dots, resonant Andreev 
transport is qualitatively different from that of non-interacting
system. Based on the
scattering matrix formalism and adopting an exact
relation for the Green's function, we have shown 
that at zero temperature the linear response conductance is enhanced due to
Kondo-Andreev resonance in the Kondo limit, while it is suppressed
in the empty site limit.  In the Coulomb blockaded region,
on the other hand, the conductance is suppressed more than the 
corresponding normal conductance even in the resonance point, 
because large charging energy suppresses Andreev reflection.

The author thanks S. Ketteman and M. Leadbeater for discussions and
comments on this manuscript. 
This work has been supported by KOSEF and in part by the visitors program of
the MPI-PKS.


%
\end{multicols}
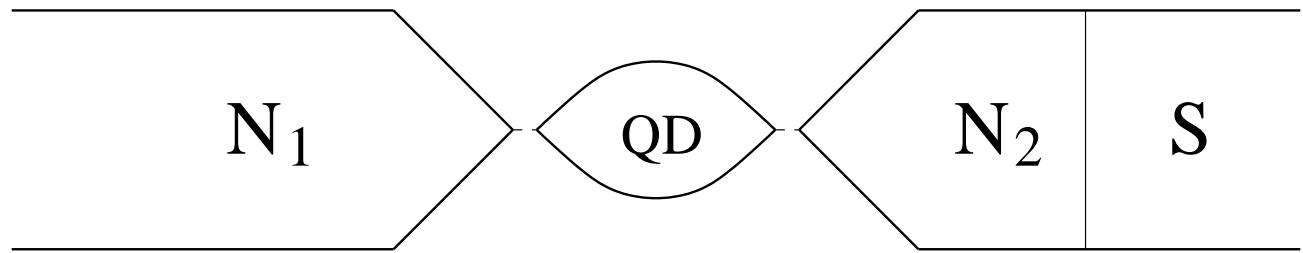
\begin{figure}
 \caption{ Schematic diagram of the quantum dot(QD) coupled to normal(N) and
           superconducting(S) leads. In the ${\rm N}_2$-S interface,
           only Andreev reflection is considered.
	   }
\end{figure}
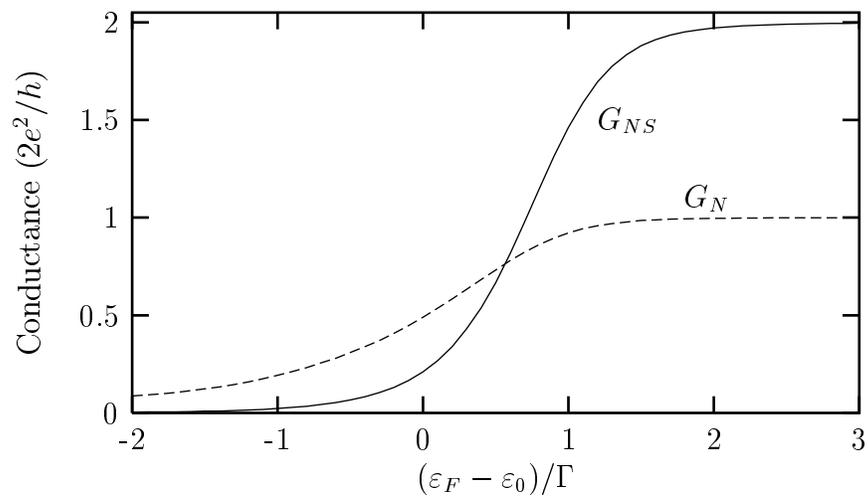
\begin{figure}
 \caption{ Conductance $G_{NS}$ and $G_N$ obtained by the Eq.(7) and 
           numerical calculation of $\varphi$ for $U=50\Gamma$ and 
           $W=200\Gamma$. }
\end{figure}
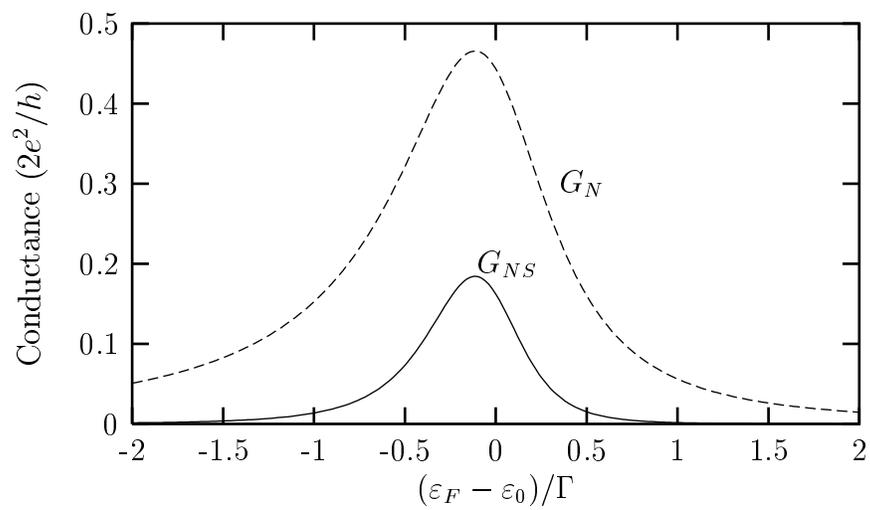
\begin{figure}
 \caption{ Conductance $G_{NS}$ and $G_N$ obtained by the Eq.(8) with (6) 
           for $U=50\Gamma$. }
\end{figure}
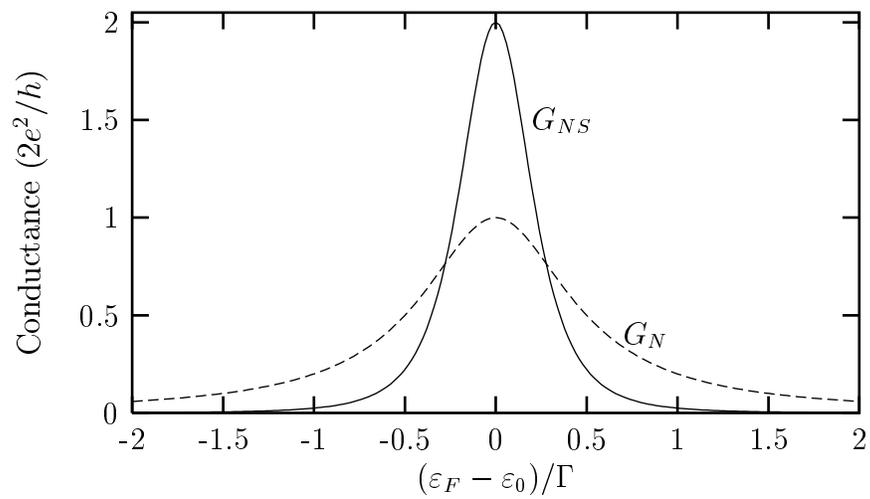
\begin{figure}
 \caption{ Conductance $G_{NS}$ and $G_N$ in the non-interacting ($U=0$)
           limit.}
\end{figure}
\end{document}